\documentclass[10pt,conference]{IEEEtran}
\IEEEoverridecommandlockouts
\usepackage{cite}
\usepackage[ruled,vlined]{algorithm2e}
\usepackage{amsmath,amssymb,amsfonts}
\usepackage{booktabs}
\usepackage{booktabs}
\usepackage{algcompatible}
\usepackage{graphicx}
\usepackage{graphicx}
\usepackage{textcomp}
\usepackage{xcolor}
\usepackage{subcaption}
\usepackage{amsmath}
\usepackage{cite}
\usepackage{amsmath,amssymb,amsfonts}
\usepackage{amsthm}
\usepackage{graphicx}
\usepackage{textcomp}
\usepackage{xcolor}
\usepackage{tikz}
\usepackage{setspace}
\usepackage{mathtools}
\usepackage{color}
\usepackage{caption}
\usepackage{subcaption}
\usepackage{lipsum}
\usepackage{soul}
\usepackage{pgfplots}
\usepackage{comment}
\usepackage{amssymb}
\usetikzlibrary{automata, positioning, arrows}
\usepackage{tabularx, booktabs}

\renewcommand{\baselinestretch}{0.96}

\newcommand{\X}{\textbf{X}}
\newcommand{\F}{\textbf{F}}
\newcommand{\G}{\textbf{G}}
\newcommand{\U}{\textbf{U}}
\newcommand{\R}{\textbf{R}}
\newcommand{\W}{\textbf{W}}

\newcolumntype{C}{>{\centering\arraybackslash}X} 
\usepackage{wrapfig}
\usepackage{ragged2e}
\usepackage{lineno}
\usepackage{float}
 
\graphicspath{ {images/} }
\definecolor{ao}{rgb}{0.0, 0.5, 0.0}
\usepackage{graphicx}
\modulolinenumbers[5]
\usepackage{adjustbox}

\def\BibTeX{{\rm B\kern-.05em{\sc i\kern-.025em b}\kern-.08em
    T\kern-.1667em\lower.7ex\hbox{E}\kern-.125emX}}
\usepackage{epstopdf}
\usepackage{url}
\usepackage{soul,color}
\usepackage{booktabs}
\usepackage{colortbl}
\usepackage[T1]{fontenc}
\usepackage{multicol}
\usepackage{amsmath}
\usepackage{authblk}
\usetikzlibrary{patterns}

\usetikzlibrary{shapes.gates.logic.US,trees,positioning,arrows,topaths}
\def\BibTeX{{\rm B\kern-.05em{\sc i\kern-.025em b}\kern-.08em
    T\kern-.1667em\lower.7ex\hbox{E}\kern-.125emX}}
\begin{document}

\title{Formal Verification for Blockchain-based Insurance Claims Processing \\
}




\author{Roshan Lal Neupane, 
Ernest Bonnah,
Bishnu Bhusal,
Kiran Neupane,
Khaza Anuarul Hoque,
Prasad Calyam \\

\small University of Missouri-Columbia, USA; {\{neupaner, kngbq, hoquek, calyamp\}@missouri.edu}, \{bhusalb, eb7z3\}@mail.missouri.edu

\thanks{This material is based upon work supported by the National Science Foundation under award numbers CNS-1950873 and CCF-1900924, and the National Security Agency under award number H98230-21-1-0260. Any opinions, findings, and conclusions or recommendations expressed in this publication are those of the author(s) and do not necessarily reflect the views of the National Science Foundation or the National Security Agency.}}


\markboth{Journal of \LaTeX\ Class Files,~Vol.~14, No.~8, August~2021}%
{Shell \MakeLowercase{\textit{et al.}}: A Sample Article Using IEEEtran.cls for IEEE Journals}


\maketitle
\begin{abstract}

Insurance claims processing involves multi-domain entities and multi-source data, along with a number of human-agent interactions. Use of Blockchain technology-based platform can significantly improve scalability and response time for processing of claims which are otherwise manually-intensive and time-consuming. However, the chaincodes involved within the processes that issue claims, approve or deny them as required, need to be formally verified to ensure secure and reliable processing of transactions in Blockchain. 
In this paper, we use a formal modeling approach to verify various processes and their underlying chaincodes relating to different stages in insurance claims processing viz., issuance, approval, denial, and flagging for fraud investigation by using linear temporal logic (LTL). 
We simulate the formalism on the chaincodes and analyze the breach of chaincodes via model checking. 
\end{abstract}


\section{Introduction}
\label{sec:1}








In the auto insurance sector, the processing of insurance claims involves gathering data from diverse sources like police, county administrators, insurance agents, and healthcare professionals~\cite{insurance-industry}. These entities work together to share crucial information, enabling insurance companies to accurately assess and handle policyholder claims. Traditional ways of performing this i.e., use of relational database along with manual processes is time-consuming and manually extensive. 

Fortunately, the widespread use of Blockchain technology-based platforms has significantly improved the efficiency and response times in processing claims. This technology enables secure sharing and management of data among untrusted organizations and entities. It offers enhanced security features such as non-repudiability, integrity, immutability, and resistance to censorship. Hyperledger Fabric~\cite{fabric}, a leading open-source blockchain platform, is widely adopted in various industrial contexts. In industries requiring authenticated users and complex data models, permissioned blockchains, supported by smart contracts/chaincodes, are crucial. Chaincodes, similar to software programs, encapsulate the business logic for creating and modifying logical assets in the ledger, and can be written in different general-purpose programming languages.

The challenges associated with this technology raises concerns about various safety and security threats, potentially leading to significant financial losses. For instance, the 2016 DAO attack led to a massive loss of 60 million USD, and a similar attack called the Parity Wallet bug resulted in a loss of 169 million USD~\cite{mehar2019understanding}. To prevent such incidents in the application of the Blockchain technology, it is crucial to employ formal verification methods to assess the underlying models of these critical systems. Formal verification specifically focuses on examining the robustness of the systems built on chaincodes~\cite{alqahtani2020formal}.

In this paper, we have implemented an approach to formally verify the functional requirements of the chaincodes that have been applied for ClaimChain~\cite{bhamidipati2021claimchain}, an exemplar consortium Blockchain-based insurance claims processing platform. To ensure security at the application and infrastructure level, we formally verify specific ClaimChain processes such as issuance, approval, etc., of insurance claims by using Linear Temporal Logic (LTL)~\cite{vasile2017time}, a modal temporal logic that have modalities referring to time. The approach establishes the implementations of the linear temporal logic of the claims processing and verifies the compliance of the system at the application and infrastructure levels. We have leveraged an LTL implementation tool, NuSMV~\cite{nusmv}, that can be used for system model verification tasks.

\begin{figure*}[ht!] 
\centering
\includegraphics[width=0.9\linewidth]{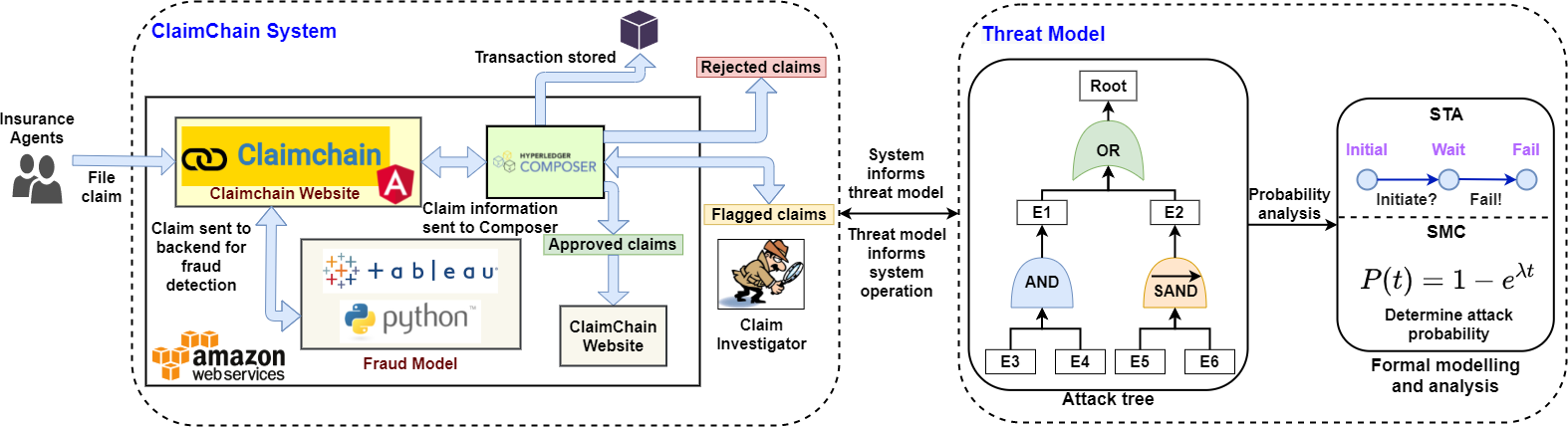}
\caption{{ClaimChain's system architecture that features threat modeling based on attack trees and fraud modeling using classifiers.
}}
\label{fig:ClaimChainarchitecture}
\vspace{-5mm}
\end{figure*}

The remainder of the paper is organized as follows:  Section~\ref{sec:2} discusses related works. Section~\ref{sec:3} gives a background on the exemplar Blockchain-based insurance processing platform. Section~\ref{sec:4} discusses formal verification of chaincodes. 
Section~\ref{sec:6} discusses the evaluation results. Section~\ref{sec:7} concludes the paper.

\section{Related Works}
\label{sec:2}
Formal verification and analysis are important to have added security to Blockchain platforms~\cite{bhargavan2016formal,matsuo2017formal}. In the literature, we have come across various methods to formally verify smart contracts (Hyperledger chaincodes in our case). Authors in~\cite{abdellatif2018formal} perform formal verification of smart contracts via users and Blockchain behavior models. They attempt to reduce smart contract vulnerabilities by verifying the smart contract breaches through a statistical model checking approach. Similarly, authors in~\cite{wang2020formal} apply formal verification on workflow policies for smart contracts in Azure Blockchain Workbench by formalizing conformance of smart contracts against state machine workflow. Authors in~\cite{sun2020formal} discuss different security vulnerabilities in Blockchain smart contracts and verify them effectively using a formal verification framework. We also found a survey of approaches to formal verification of the Blockchain smart contracts~\cite{murray2019survey}.

Our novelty lies in analyzing processes in an exemplar Blockchain-based insurance claims processing platform where we formally verify the operations (such as e.g., issuance, approval), and the endorsement policy within Blockchain using LTL.

\begin{figure}[h] 
    \centering
    \vspace{-2mm}
    \includegraphics[width=\linewidth]{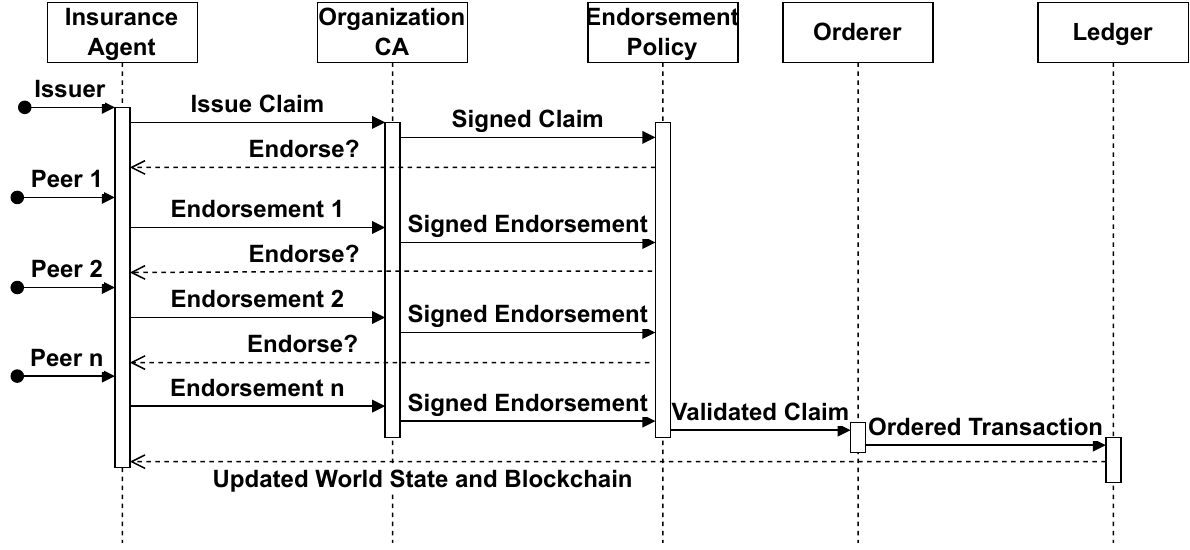}
    \caption{{Transaction issuance process within the functions involved in the Hyperledger Chaincode.}}
    \label{fig:issueClaim}
    \vspace{-5mm}
\end{figure}

\section{Background}
\label{sec:3}
In this section, we provide a background on ClaimChain, which is an exemplar platform for Blockchain-based insurance claims processing. For our study purposes, we investigate the soundness of ClaimChain processes by leveraging formal verification of the chaincodes developed for claim issuance, claim approval, claim denial, and claim flagging. 

\subsection{ClaimChain Architecture}
Figure~\ref{fig:ClaimChainarchitecture} depicts the overall ClaimChain platform architecture. ClaimChain is developed on Hyperledger Fabric~\cite{fabric}, a Linux Foundation project for developing Blockchain platforms. 
All the peers/organizations i.e., participating insurance companies are connected to Hyperledger Fabric network. Agents of such organizations and other multi-domain entities use our UI developed in Angular~\cite{angular} to initiate a transaction i.e., an insurance claim. When they submit a claim, the frontend makes an API call to trigger the issue chaincode. The user transactions are validated, and inserted into a block and dispersed within the shared Blockchain. 


In the ClaimChain architecture, there are two critical components: the threat model, which bolsters infrastructure-level security, and the fraud model, designed to fortify application-level security. To conduct our threat modeling, we apply the attack tree formalism~\cite{attacktree} to pinpoint various scenarios where data integrity attacks could occur. We determine the likelihood of these attacks happening at the infrastructure level through a thorough analysis of the ClaimChain-specific attack tree. Concurrently, ClaimChain employs fraud model to spot deceptive claims by monitoring data generated during the handling of user queries in application-level operations. The fraud modeling leverages supervised machine learning techniques to scrutinize for fraudulent activities, using red flags identified by the National Insurance Crime Bureau (NICB)~\cite{nicb} to accurately and reliably identify instances of fraud.

\begin{figure}[h] 
    \centering
    \vspace{-3mm}
    \includegraphics[width=\linewidth]{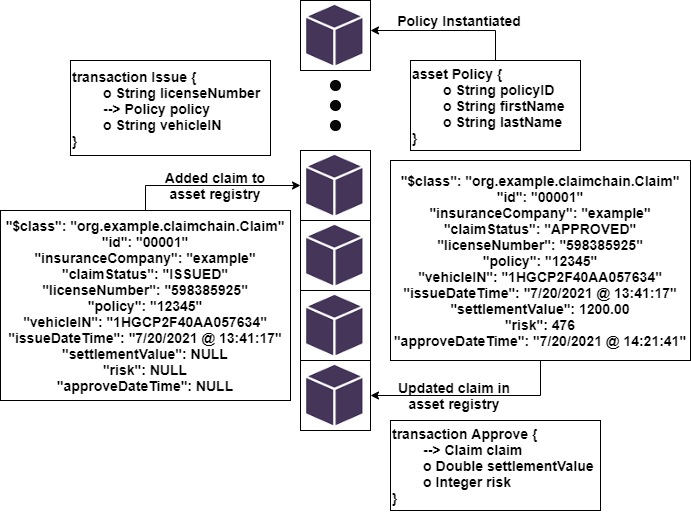}
    \caption{{Transaction issuance and approval processes within the functions involved in the Hyperledger Chaincode.}}
    \label{fig:ClaimChain}
\vspace{-5mm}
\end{figure}

\begin{figure*}[t] 
    \centering
    \vspace{-2mm}
    \includegraphics[width=.7\linewidth]{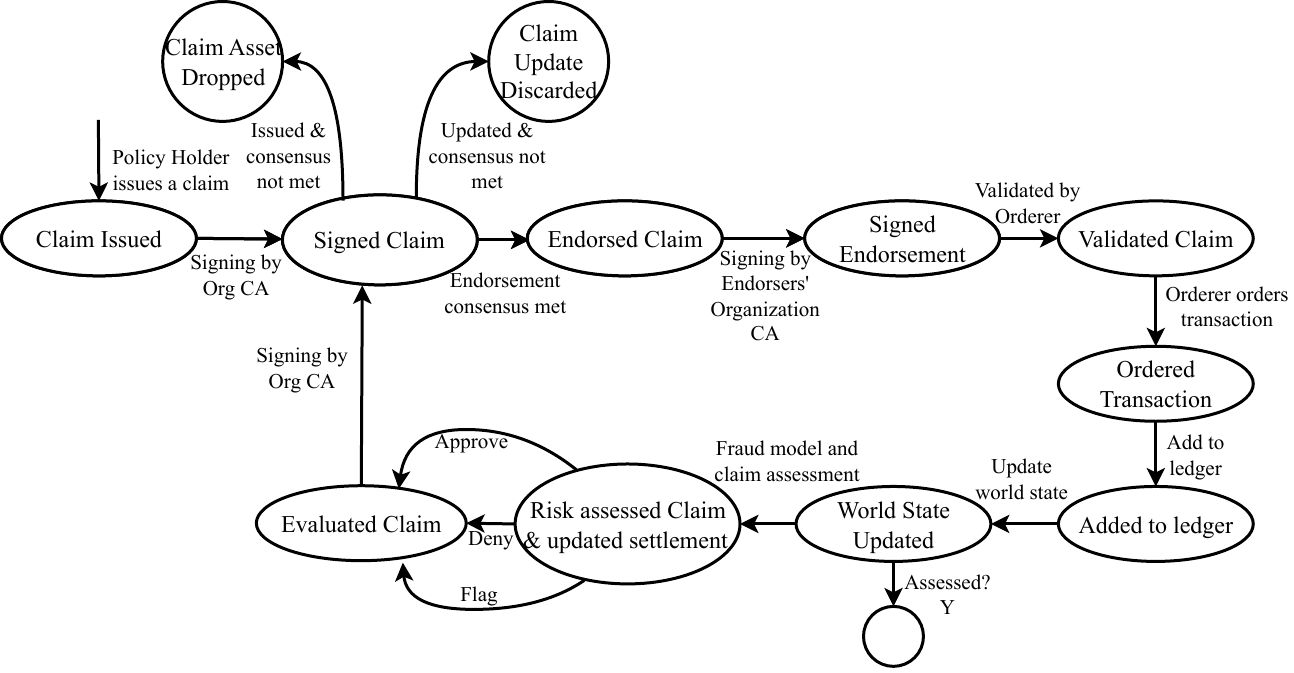}
    \caption{ClaimChain model process flow for model checking the trustworthiness of the system}
    \label{fig:FSM}
    \vspace{-5mm}
\end{figure*}

\subsection{ClaimChain Processes}
Figure \ref{fig:ClaimChain} illustrates the life cycle of a ClaimChain claim asset, spanning from policy instantiation to claim approval. A more detailed examination of the "Issue" operation is presented in Figure \ref{fig:issueClaim}. In this process, an insurance agent receives a claim from a policyholder, triggers the `issue claim' smart contract, and attaches their certificate authority (CA). Crucial claim details, such as the policyholder's license number, policy ID, and vehicle identification number, are incorporated into the claim asset. The peer then evaluates the transaction via simulation, endorses it by attaching their CA, and the client application proceeds to obtain endorsements from Endorsing peers in the channel, adhering to the specified endorsement policy. Subsequently, the peer initiates a request to the orderer for recording the transaction on the Blockchain. After the transaction's order is determined, it is packaged into a block and disseminated to network peers for inclusion in their records. The claim asset is only marked as issued and added to the World state after the transaction is securely recorded. During fraud detection, the fraud model interrogates the World state to verify the presence of duplicate claims and scrutinizes the claim information for any NICB-identified red flags that may indicate fraud.


If the claim is approved,
the agent triggers the `approve claim' smart contract and appends their Certificate Authority (CA). This contract, using the provided claim identifier, selects a corresponding claim from the World state, incorporates risk scores and settlement values, and officially marks it as approved by the organization. Once again, the peer validates the transaction through simulation and provides its endorsement by appending its CA. The client application continues to gather endorsements from the Endorsing peers within the channel, following the criteria specified by the endorsement policy. Subsequently, the peer submits a request to the Orderer for the transaction's inclusion on the Blockchain. After the order of the transaction is determined, it is packaged into a block and distributed to peers throughout the network for integration into their records. Only after the transaction is securely recorded is the claim designated as approved and updated within the World state. Once a claim asset has received approval, it becomes immutable and cannot be modified, though it remains stored in the World state for future reference.

The process for evaluating claims that result in 'denial' or 'flagging' follows a similar sequence. Furthermore, the 'cancel policy' transaction is employed to conclude a policy with the organization. Similarly, after a policy has been canceled, it becomes read-only, persisting in the World state without the possibility of further interaction.

\section{Formal Verification of ClaimChain Processes}
\label{sec:4}

Figure~\ref{fig:FSM} represents the processes within the ClaimChain platform from the point where a claim is issued and added to the world state, to the point where claim decision/evaluation is made and updated in the ledger and then subsequently updated in the world state. To reason events in ClaimChain transactions, such as issuance of claims, approvals, denials, etc., we use Linear Temporal Logic (LTL) which is a formal way of representing properties for the Chaincodes involved in the events. The general syntax for LTL is:

\begin{equation*}
\footnotesize
\label{eq:ltlsyntax}
  \begin{aligned}
    &\phi ::= \top | \bot | p | (\neg \phi) | (\phi \land \phi) | 
    (\phi \lor \phi) | (\phi \rightarrow \phi) | (\X\phi) |  \\ 
    &~~~~~~~~(\F\phi) | (\G\phi) | (\phi \U \phi) | (\phi \W \phi) | (\phi \R \phi)
 \end{aligned}
\normalsize
\vspace{-2mm}
\end{equation*}


\noindent where, 
\begin{itemize}
    \item $\top$: Represents a state that is always true, indicating that a given property holds at all times. 
    \item $\bot$: Represents a state that is always false, indicating that a given property never holds at any time.
    \item $p$: Represents a propositional variable or atomic proposition, which can take on true or false values. These variables are used to describe the state of the system.
    \item $\neg \varphi$: The negation of a formula $\varphi$ is true in a state if $\varphi$ is false in that state.
    \item $\varphi \land \psi$: The conjunction of formulas $\varphi$ and $\psi$ is true in a state if both $\varphi$ and $\psi$ are true in that state.
    \item $\varphi \lor \psi$: The disjunction of formulas $\varphi$ and $\psi$ is true in a state if either $\varphi$ or $\psi$ is true in that state.
    \item $\X \varphi$: The next operator ensures that a formula $\varphi$ is true in the next state.
    \item $\F \varphi$: The eventually operator requires that a formula $\varphi$ becomes true at some point in the future.
    \item $\G \varphi$: The globally operator states that a formula $\varphi$ must be true at all future states.
    \item $\varphi \U \psi$: The until operator specifies that a formula $\varphi$ must be true until a formula $\psi$ becomes true, and then $\psi$ must remain true.
    \item $\varphi \W \psi$: Represents the ``weak until'' operator. It's similar to the ``until'' operator but allows $\varphi$ to be true even if $\psi$ is never true.
    \item $\varphi \R \psi$: Represents the ``release'' operator. It means that $\varphi$ must hold true until $\psi$ becomes true, and then it allows $\varphi$ to become true as well. It's a combination of both the ``until'' and ``weak until'' operators.
\end{itemize}
Below, we show  LTL specifications and their natural language translations:

\begin{itemize}
    \item \textbf{LTL Specification 1 ($\varphi_1$)}:
    When the stage is $endorsed$ it is required that at some point in the future (\F), the claim status will become $approved$, $denied$, or $flagged$.
    \begin{equation}
    \footnotesize
        \begin{multlined}
       \text{stage} = \text{endorsed} \rightarrow \text{\F(claim\_status = approved) } |  \\
         \text{\F(claim\_status = denied) } | \text{ \F(claim\_status = flagged)} \nonumber
    \end{multlined}
    \normalsize
    \end{equation}

    \item \textbf{LTL Specification 2 ($\varphi_2$)}: 
    If the stage is $issued$, it is required that at some point in the future (\F), the stage will become $claim\_asset\_dropped$, $claim\_updated\_discarded$, or $evaluated\_world\_state\_updated$.
        \begin{equation}
        \footnotesize
        \begin{multlined}
            \text{stage} = \text{issued}  \rightarrow \text{\F(stage = claim\_asset\_dropped) }  | \\
             \text{\F(stage = claim\_updated\_discarded) }   | \\
             \text{\F(stage = evaluated\_world\_state\_updated)}\nonumber 
         \end{multlined}
         \normalsize
        \end{equation}

    \item \textbf{LTL Specification 3 ($\varphi_3$)}:
    If the stage is $issued$, then it is guaranteed that, at some point in the future, the stage will enter the $endorsed$ state.
    \begin{equation}
    \footnotesize
        \begin{multlined}
            \text{stage} = \text{issued} \rightarrow \text{\F(stage = endorsed)} \nonumber
        \end{multlined}
        \normalsize
    \end{equation}

    \item \textbf{LTL Specification 4 ($\varphi_4$)}:
    When the stage is $signed$ and it is not followed by the $endorsed$ stage, it must continuously remain in the $claim\_asset\_dropped$ state.
    \begin{equation}
    \footnotesize
        \begin{multlined}
        \text{stage} = \text{signed } \& \text{ stage} \neq \text{endorsed} \\ \rightarrow \text{\G(stage = claim\_asset\_dropped)} \nonumber
        \end{multlined}
        \normalsize
    \end{equation}

    \item \textbf{LTL Specification 5 ($\varphi_5$)}:
    If the stage is $issued$, then it is guaranteed that, at some point in the future, the stage will enter the $evaluated$ state.
    \begin{equation}
    \footnotesize
        \begin{multlined}
            stage = \text{issued} \rightarrow \text{\F(stage = evaluated)} \nonumber
        \end{multlined}
        \normalsize
    \end{equation}

\end{itemize}

\begin{figure}[h] 
    \centering
    \includegraphics[width=0.75\linewidth]{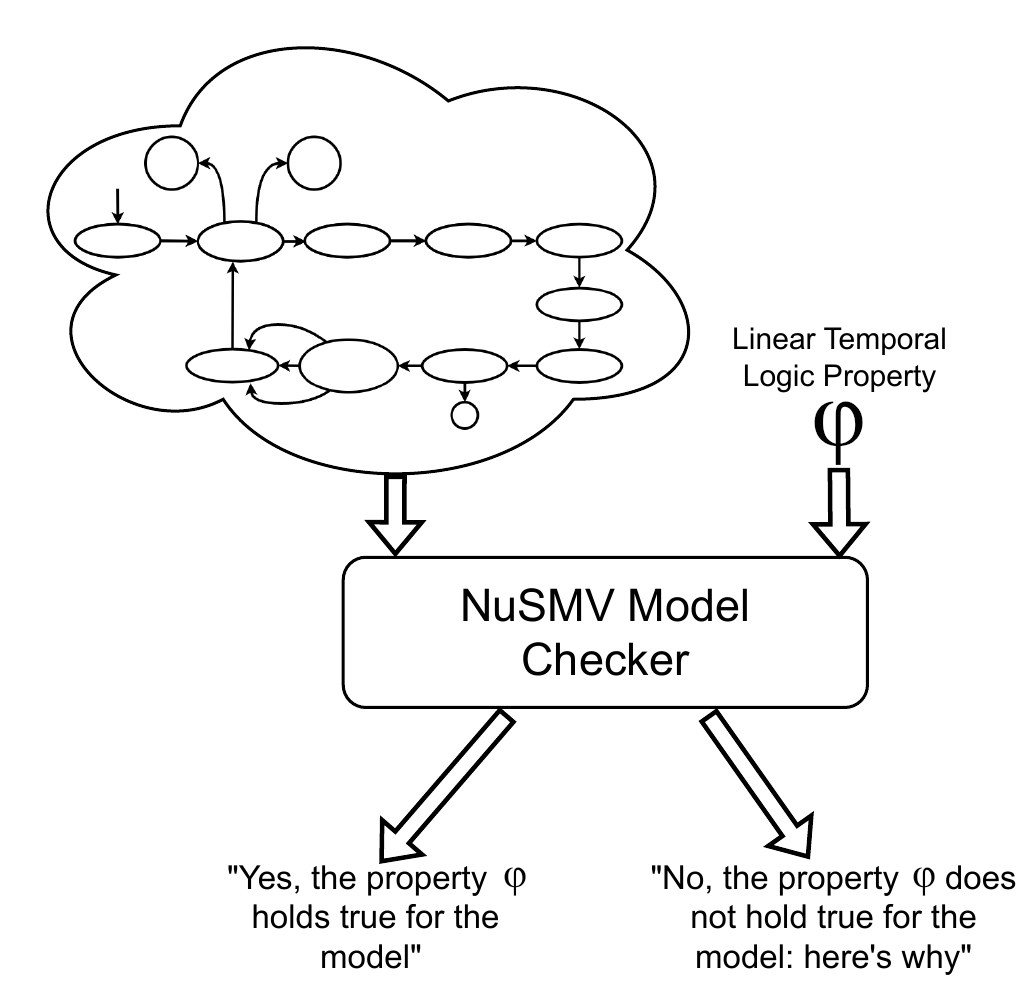}
    \caption{{NuSMV model checking for process flow of the ClaimChain model}}
    \label{fig:NuSMV}
\vspace{-4mm}
\end{figure}

\section{Evaluation Results}
\label{sec:6}
We simulate the specifications described in Section~\ref{sec:4} using the model-checking tool NuSMV. The simulation and experiments are carried out on a 1.4 GHz Quad-Core Intel Core i5 equipped with 8 GB of RAM. Figure~\ref{fig:NuSMV} depicts that the model checking involves the utilization of a model checker (NuSMV in our case), which accepts a model that represents a finite state abstraction (as depicted by Figure~\ref{fig:FSM} of the platform), along with a statement regarding the platform's behavior expressed in temporal logic (as discussed in the specifications). Subsequently, the NuSMV assesses whether the statement holds true for the model or not, and in the event of the statement being false, most practical model checkers will offer a counterexample. 

\begin{figure}[h] 
    \centering
    \includegraphics[width=0.95\linewidth]{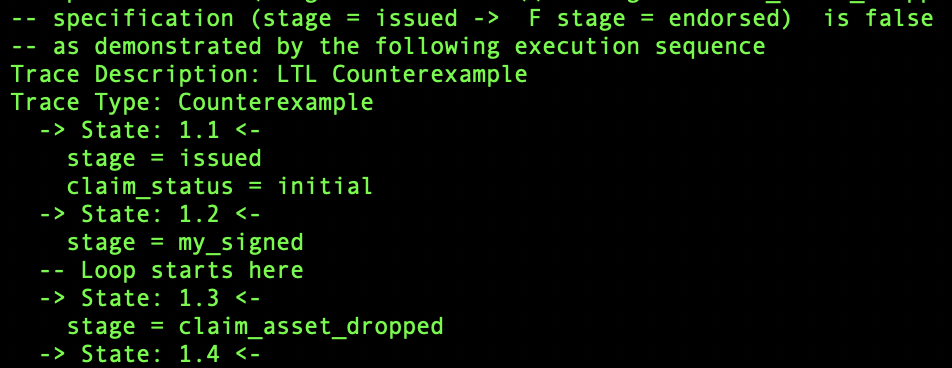}
    \caption{{Counterexample for the unsatisfied specification $\varphi_3$}}
    \label{fig:spec3}
    \vspace{-3mm}
\end{figure}


\begin{table}[h]
    \centering
    \caption{Experimental results for model checking LTL formulae discussed in Section~\ref{sec:4}}
    \begin{tabular}{|c|c|c|}
    \hline
        Specifications & Verdict & Time (in seconds) \\ \hline
        $\varphi_1$ & $\top$ & 0.01 \\ \hline
        $\varphi_2$ & $\top$ & 0.01 \\ \hline
        $\varphi_3$ & $\bot$ & 0.02 \\ \hline
        $\varphi_4$ & $\top$ & 0.01 \\ \hline
        $\varphi_5$ & $\bot$ & 0.02 \\ \hline
    \end{tabular}
\label{tab:sat}
\vspace{-3mm}
\end{table}

Table~\ref{tab:sat} shows the model checking for the specifications and their verdict. Essentially, $\top$ means the specification is satisfied by the model, and $\bot$ means the specification is not satisfied by the model. The platform satisfies the LTL specifications $\varphi_1$, $\varphi_2$ and $\varphi_4$. However, the rest of the two specifications, $\varphi_3$ and $\varphi_5$ were unsatisfied. For a more in-depth explanation of our observation, let's take an example of Specification 3 ($\varphi_3$) - The assertion made in Specification 3 implies a deterministic relationship between the issuance of a stage and its eventual endorsement. However, our examination reveals scenarios, particularly within blockchain contexts, where the straightforward fulfillment of this specification encounters limitations. In such instances, the failure to achieve consensus among the majority parties within the blockchain ecosystem can impede the expected progression of a claim from the ‘issued’ to the ‘endorsed’ state. 

We also recorded the time taken to check the satisfiability of the specifications above. The model checking tool also provides counterexample to unsatisfied specifications. Figure~\ref{fig:spec3} shows such an example for the LTL specification 3 ($\varphi_3$). 

\section{Conclusion and Future Work}
\label{sec:7}

In this paper, we have conducted a comprehensive analysis of an exemplar Blockchain-based insurance claims processing platform viz., ClaimChain, focusing on formal verification of its underlying processes that includes claim issuance, approval, denial and flagging for further investigation
In our research with ClaimChain, we converted chaincodes into a Finite State Machine (FSM) to model interactions and transitions within ClaimChain. They were then verified for correctness using the NuSMV tool. The functional requirements of insurance claims processes were directly integrated into the tool using Linear Temporal Logic (LTL) properties. 

As part of future work, one can introduce more complex specifications that can cover other aspects of the insurance claims platform such as the policy instantiation, transaction on blocks, and other relevant components of Blockchain.

\bibliographystyle{IEEEtran}
\bibliography{ref.bib}

\end{document}